\documentclass[preprint, 12pt]{elsarticle}
\usepackage{lineno,hyperref}
\hypersetup{colorlinks = true}
\journal{Physics of Fluids}
\biboptions{authoryear}
\usepackage{amsmath}
\usepackage{caption}
\usepackage{subcaption}
\usepackage{graphicx}
\usepackage{upgreek}
\graphicspath{ {img/} }
\usepackage{geometry}% by courtesy of Mico
\usepackage{array}

\usepackage{setspace}
 
\usepackage{upgreek} %upright greek letters in math mode
\usepackage[T1]{fontenc}
\usepackage{kpfonts}
\begin{document}
\begin{frontmatter}
	\title{Enhancing respiratory comfort with fan respirators: computational analysis of carbon dioxide reduction, temperature regulation, and humidity control}

	\author[label1]{Hana Salati}
	\author[label1]{Patrick Warfield-McAlpine}
	\author[label3]{David F Fletcher}
	\author[label1]{Kiao Inthavong \corref{cor1}}

	\cortext[cor1]{Corresponding Author: kiao.inthavong@rmit.edu.au}
	\address[label1]{Mechanical \& Automotive Engineering, School of Engineering, RMIT University, Bundoora, VIC, 3083, Australia}
    \address[label3]{School of Chemical and Biomolecular Engineering, The University of Sydney, Sydney, NSW, 2006 Australia}

{\setstretch{1.0}	
\begin{abstract}
Respirators provide protection from inhalation exposure to dangerous substances, such as chemicals and infectious particles, including SARS-Covid-laden droplets and aerosols. However, they are prone to exposure to stale air as the masks create a microclimate influenced by the exhaled air. As a result, exhaled air from the lungs accumulating in the mask produce a warm and humid environment that has a high concentration of carbon dioxide (CO$_2$), unsuitable for re-inhalation. Fans are a favourable option for respirators to ventilate the mask and remove the stale air. This study utilized computational fluid dynamics simulation consisting of a hybrid Reynolds-averaged Navier-Stokes (RANS)-large eddy simulation (LES) turbulence method to compare the inhalation flow properties for different fan locations (bottom, top, and side) with regular respirator breathing. Three mask positions, top, side, and bottom, were evaluated under two breathing cycles (approximately 9.65s of breathing time). The results demonstrated that adding a fan respirator significantly decreased internal mask temperature, humidity, and CO$_2$ concentration. The average CO$_2$ concentration decreased by 87\%, 67\% and 73\% for locations bottom, top and side respectively. Whilst the top and side fan locations enhanced the removal of the exhaled gas mixture, the bottom-fan respirator was more efficient in removing the nostril jet gas mixture and therefore provided the least barrier to respiratory function. The results provide valuable insights into the benefits of fan respirators for long-term use for reducing CO$_2$ concentration, mask temperature, and humidity, improving wearer safety and comfort in hazardous environments, especially during the COVID-19 pandemic.
\end{abstract}
}	
\end{frontmatter}

%\linenumbers
\section{Introduction}
Although over two years have passed since the start of the COVID-19 pandemic, masks and respirators are still mandatory in some indoor settings such as public transport and hospitals. Face mask requirements vary between different countries and transmission peak times. Sense of nasal stuffiness, dry nose, overheating, and nasal obstruction are the feelings experienced by healthcare workers \citep{purushothaman2021effects}. Exposure to CO$_2$ results in various health effects, including increased respiratory rate, brain blood flow, metabolic stress, headache, dizziness, confusion, and dyspnea   \citep{atangana2020facemasks,azuma2018effects,raub2000carbon}. \cite{salati2021n95} reported that a respirator could cause excessive CO$_2$ inhalation by approximately $7 \times$ greater per breath compared with normal breathing. Fan-powered masks have been proposed to resolve respiratory-related complications and provide breathing comfort for an extended period.

Previous studies have investigated  mask-related topics including filtration \citep{dbouk2020respiratory,verma2020visualizing, arumuru2020experimental}, mask leakage \citep{solano2022perimeter, kang2021particle}, and filter material \citep{xi2022inspiratory,hossain2020recharging}. Limited studies focused on how fresh air can be supplied to the dead-space of the mask for the user and how exhalation valves change the flow characteristics inside the mask. \cite{article} proposed a battery-powered active respirator design that uses an exhalation valve on the side of an N95 respirator to provide fresh air with every breath. The CO$_2$ concentration for the conventional N95 and modified N95 were compared with normal breathing (control). The relative CO$_2$ concentration of the control case was 100$\%$, and modified N95 reduced it to 16.67$\%$.
\cite{zhang2016improved} measured the temperature and CO$_2$ concentration within the dead-space of an N95 mask using a fan installed at the middle front region of the mask. CFD simulations and experimental measurements showed that adding the ventilation fan reduced the dead-space temperature by 2$^\circ$C. \cite{staymates2020flow} analyzed the flow characteristics of an N95 respirator with and without an exhalation valve using Schlieren imaging. The results demonstrated that N95 respirators with exhalation valves are not suitable for reducing the proliferation of infectious diseases. Despite the increasing use of fan respirators in various hazardous settings, there is a lack of comprehensive studies investigating the effects of different fan locations on inhalation flow properties, highlighting the need for further research in this area. 

This paper aims to understand the breathing airflow characteristics inside an N95 respirator with different fan locations (bottom, top, and side of the respirator). This study compares the re-inhalation air quality (air CO$_2$ and H$_2$O concentration and temperature) from different fan locations with N95 respirator breathing (without a fan) and normal breathing (without a respirator). To achieve these objectives, we applied a hybrid RANS-LES turbulence model to investigate the fluid characteristics of the nasal airflow jets inside a fan-powered N95 respirator during two consecutive breathing cycles (8 seconds).

\section{Method}
\subsection{Geometry model}
A 3D model of the N95 respirator was imported into the modelling CAD software (Ansys SpaceClaim), then positioned and attached to the human face. The respirator was a non-fully sealed mask representing the natural leakage of a respirator \citep{salati2021n95}. A gap was considered between the respirator and the face, ranging from 1mm to 2mm due to the curvature of the face. This study investigated the effects of adding a fan to existing N95 respirators used in clinical practice. As such masks are designed to be lightweight and disposable, they do not contain the necessary materials to provide an airtight seal between the mask and the wearer’s face. Subsequently, a leakage threshold was applied to best represent real-world applications. Further investigations into more advanced sealable respirators may be investigated in future studies.

\begin{figure}[h!]
	\centering
	\includegraphics[width=1\linewidth]{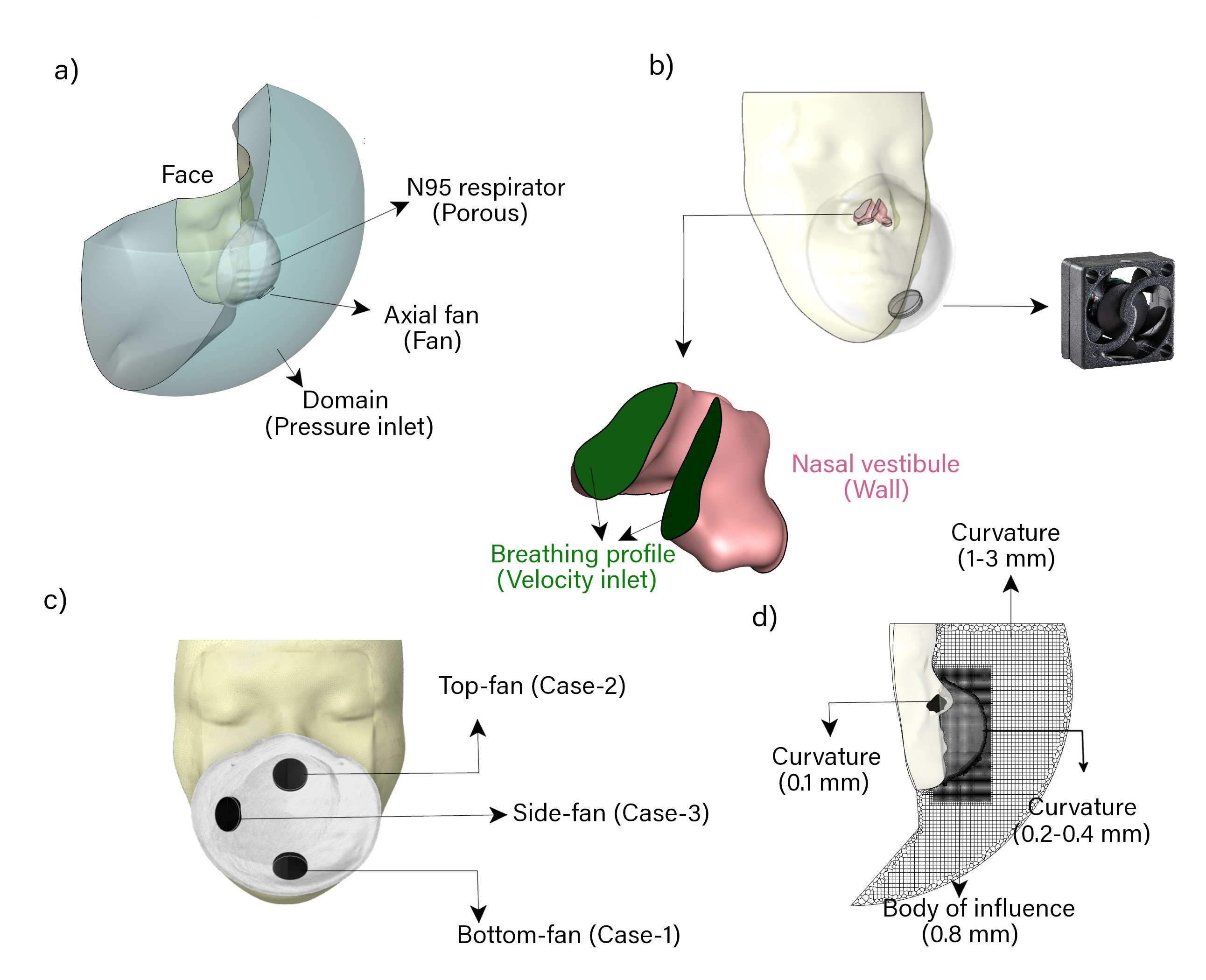}
	\caption{(a) Computational domain and applied boundary conditions for ambient, fan, and respirator. (b) Nasal vestibule and breathing boundary condition surfaces. (c) Different fan locations include the bottom-fan, side-fan, and top-fan respirator. (d) Cross-sectional view of the volume mesh and sizing. }
	\label{fig:bc}
\end{figure}

The nasal vestibules from a healthy human subject were taken from a CT scan and imported to 3D Slicer\textregistered \space segmentation software. A micro fan (Pelonis Technologies, AGA178) was represented with a cylindrical volume of $1.25 \times 10^{-6}$ m$^3$ (diameter of 20 mm) on the respirator in three different locations top, bottom, and side as shown in Figure~\ref{fig:bc} (geometrical details of the fan were neglected). 

\subsection{Boundary conditions and meshing}
An external domain outside of the nose was created in front of the face where the exterior surface of the domain was set to atmospheric pressure. The respiration was initiated by defining a mass flow rate at the flow exit located at the internal region of each nostril (i.e., nasal vestibule). The respiratory cycle was simplified to pure sine waves based on the measured physiology data from \cite{benchetrit1989individuality} and used in \cite{Calmet2020}, allowing a simple method for describing the tidal volume, breathing periods, and periodicity:

\begin{equation}
    \dot{m}=A \sin{\left(\frac{\pi(t-C)}{B}\right)}  
\end{equation}

\begin{figure}[h!]
	\centering
	\includegraphics[width=1\linewidth]{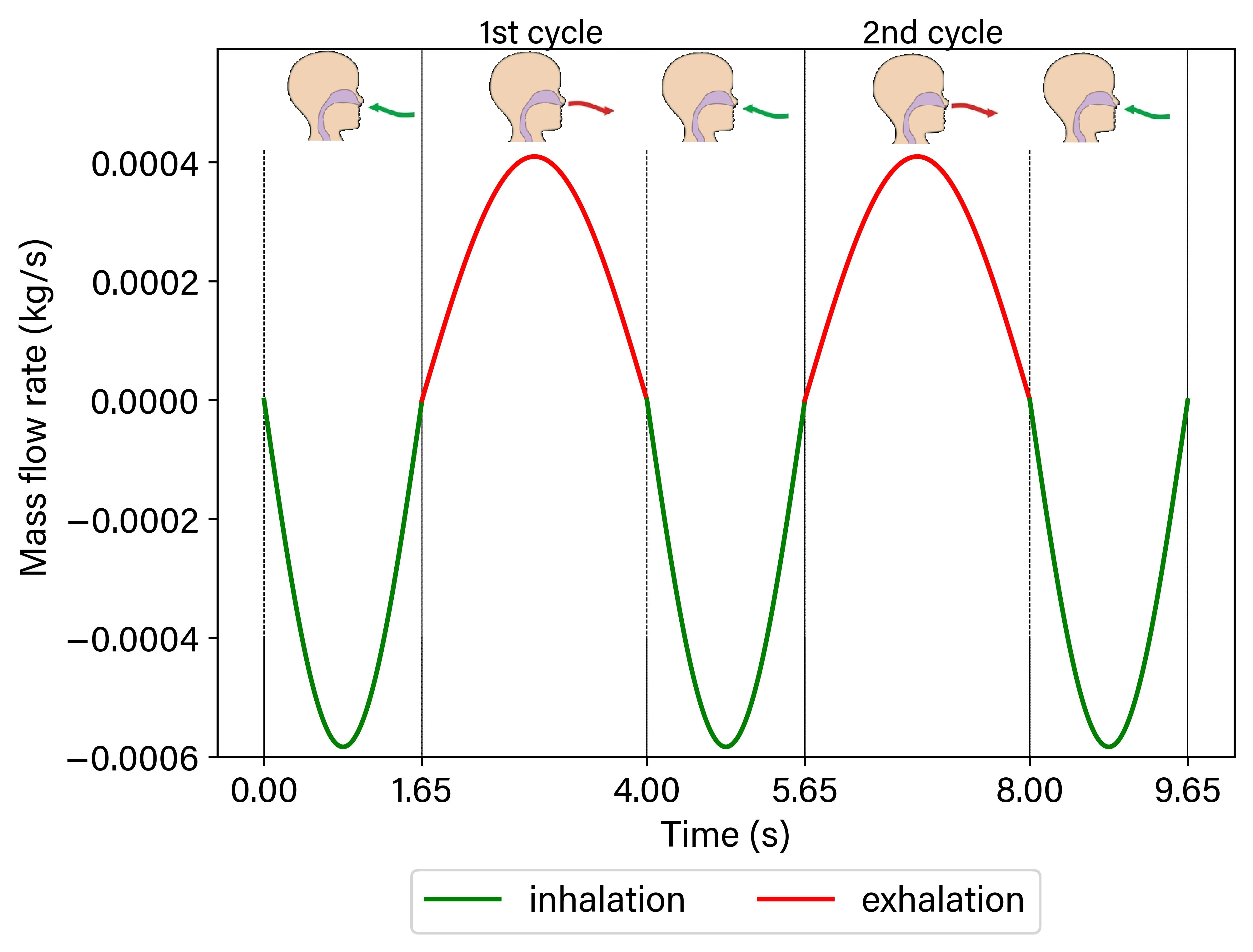}
	\caption{Breathing cycle mass flow rate profile assigned to the nostrils. The mass flow rate was divided equally between the nostrils and converted to the velocity profile. The simulation was performed for two complete respiration cycles with an additional inhalation phase which was discarded to avoid the startup effects on the results.}
	\label{fig:time}
\end{figure}

For inhalation the amplitude was $ A = 5.832 \times 10^{-4}$ ~kg/s; period $B =1.65$~s; while the periodicity for multiple breathing cycles was $C = 4(n-1)$~s, where $n$ is the respiration cycle number. For exhalation $ A = 4.0945 \times 10^{-4}$ ~kg/s; $B =2.35$~s; and $C = (1.65 + 4(n-1))$~s. This expression produces a tidal volume of  500~mL, where the inhalation period is 1.65~s and the exhalation period is 2.35~s. The mass flow rate was divided between the nostril outlets evenly, although in reality, most nasal cavities exhibit some asymmetric flow distribution \citep{tan2012numerical,wang2016investigation,kahana2016measuring}. The solution was initialized at time $t=0$~s with steady-state settings of ambient conditions and then continued for 9.65 seconds (Figure \ref{fig:time}). 
\clearpage

The simulation time included three inhalation and two exhalation periods. The first inhalation phase was excluded to avoid startup effects from the analysis and to ensure the respiration results represented continuous breathing. The ambient air temperature, relative humidity, and CO$_2$ concentration were set to 20$^\circ$C, 30\%, and 385 ppm, respectively. The exhaled air temperature, relative humidity, and CO$_2$ concentrations were assumed to be 36$^\circ$C, 100\%, and 36000 ppm \citep{salati2021n95}.

The model was meshed following the method used in \cite{van2021pressure} where the LES component of the SBES model had a sufficiently fine mesh to resolve the large eddies. The model was meshed with poly-hexcore elements using Ansys Fluent 2021R1. A minimum surface sizing of 0.1 mm and 0.4 mm was applied to all the regions. A body of influence was applied to the front of the mask. Local mesh element refinements for regions of rapid flow changes were created through `body of influence' mesh zones with a sizing of 0.8~mm was applied. Five prism layers were applied with a last ratio of 28 $\%$ and a first layer height of 0.05~mm  (Fig.~\ref{fig:bc}d) ensuring the normalized wall distance y+ < 1 on all walls. The strict requirements of normalized wall distances x+ and z+ were not required for the SBES model. We selected an iterative meshing strategy to have a sufficiently fine mesh to resolve the energy-containing eddies. The first step meshed the model using 1.5 million poly-hexcore elements with five inflation layers on all walls. Then the model was run in a pre-cursor steady state simulation, and the ratio of the smallest resolved length scale to the turbulence integral length scale was checked to see if it resolved 80\% of the turbulence kinetic energy. Finally, we refined the region of interest using a body of influence, with the final mesh consisting of 2.4 million cells. 

The N95 mask surface was set as a porous zone with a porosity of 0.88, and a viscous resistance coefficient of $1.12 \times 10^{10}$~m$^{-2}$ based on \cite{zhang2016investigation}, and \cite{salati2021n95}. The flow rate of the fan was 0.8 CFM. A fan boundary condition was used, with a 5 Pa pressure jump applied to reproduce the fan flow rate. The pressure jump was regulated to reach the targeted flow rate of this specific fan. The wall temperature was set to 34$^\circ$C in this region, and the wall thickness was assumed to be 6~mm \citep{senanayake2021impact,lindemann2002nasal,Lindemann2004,salati2021n95,keck2000humidity}.

\subsection{Numerical setup}
This study used the Stress-Blended Eddy Simulation (SBES) turbulence model. SBES is a hybrid RANS-LES model that applies a Large Eddy Simulation (LES) in the bulk flow region to resolve large eddies and an $k-\omega-$SST model was used in Reynolds-Averaged Navier Stokes (RANS) region near the wall \citep{ph15101259,bradshaw2022new,menter2018stress,wickramarachchi2023comparison}. Recent studies have adopted the SBES model to simulate the respiratory flow field \citep{ph15101259,bradshaw2022new}. Similarly, in our previous study, we verified the  SBES model to be reliable in depicting airflow through a nasal cavity airway by comparing the pressure variation at different regions of the airway with experimental measurements in a 3D-printed model in \cite{van2021pressure}. The study demonstrated that applying best practices including sensitivity analyses for mesh size and time steps ensured the accuracy and reliability of the numerical simulations.

The incompressible flow equations describing the conservation for mass and momentum are expressed as:
\begin{align}
    \frac{\partial}{\partial x_i} \left( \bar{u}_i\right) &= 0 \\
    \frac{\partial}{\partial t}\left(\rho \bar{u}_i\right) + \frac{\partial}{\partial x_j}  \left(\rho \bar{u}_i \bar{u}_j \right) &=  -\frac{\partial \bar{p}}{\partial x_i} + \frac{\partial \sigma_{ij}}{\partial x_j} + \frac{\partial \tau_{ij}}{\partial x_j} 
\end{align}
where $ u_i $ is the flow velocity vector,  $ \rho $ is the fluid density and $ p $ is the pressure, and $\sigma_{ij}$  is the stress tensor due to molecular viscosity. The overbar $ \bar{\varphi} $ on the scalar quantity $ \varphi $ denotes a Reynolds-averaging operation in the RANS formulation and a spatio-temporal filtering operation in the LES formulation. The turbulence stress tensor $ \tau_{ij} \equiv \rho \left( \overline{u_i u_j} - \bar{u}_i \bar{u}_j \right) $ for the SBES formulation is defined to blend between the Reynolds stress tensor $ \tau_{ij}^\text{RANS} $ for the RANS formulation and the subgrid-scale stress tensor $ \tau_{ij}^\text{LES} $ for the LES formulation, according to the blending function:
\begin{equation}
    \tau_{ij} = f_s \tau_{ij}^\text{RANS} + \left( 1 - f_s \right) \tau_{ij}^\text{LES}
\end{equation}
where $ 0 \le f_s \le 1 $ is the shielding function. 

The time-step size chosen aimed to resolve the energy-containing eddies. An estimate of the time resolution of the model was calculated based on the time scale. To maintain stability whilst minimising simulation time a variable time-step was employed to ensure a local Courant–Friedrichs–Lewy (CFL) number < 1. At peak inhalation and exhalation, a minimum time step of 1 $\times 10^-5$ was used. As the flow rate decrease the time step was increased to 1 $\times 10^-4$ for computational efficiency.

The SIMPLE scheme was selected for pressure-velocity coupling. The convective terms of the momentum equations were resolved using second-order bounded schemes. The gradients were calculated using the least-squares cell-based methodology and time stepping was discretised using a second-order bounded implicit scheme. Default under-relaxation factors were employed. Residual monitors for energy were set to 1 $\times 10^-6$ and the remainder set to 1 $\times 10^-4$.

\section{Results}
\begin{figure}[h!]
	\centering
	\includegraphics[width=1.1\linewidth]{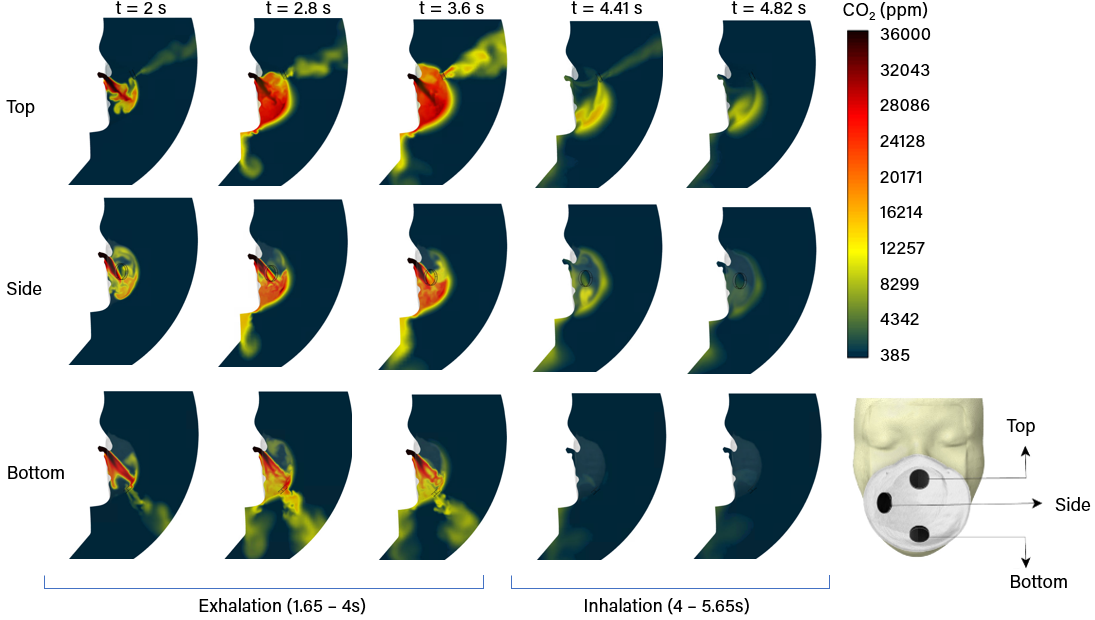}
	\caption{CO$_2$ distribution at the mid-sagittal plane of the right nasal passage for different fan locations.}
	\label{fig:co2-cont}
\end{figure}

The ambient and exhalation CO$_2$ level is 385 ppm and 36,000 ppm, respectively (the exhalation CO$_2$ values is 93 times more than the ambient level). Exposure to a CO$_2$ concentration of 10,000 ppm for 30 minutes or more in healthy adults results in respiratory acidosis, and excessive CO$_2$ can cause increased respiratory rate and metabolic stress. The CO$_2$ distribution in the mid-sagittal plane of the right nasal passage during the first respiration cycle ($t=1.65$ to 4.00 s) due to different fan locations is shown in Figure~\ref{fig:co2-cont}. There were three fan locations (top, side, and bottom) which are shown in Figure \ref{fig:bc}c.

At $t= 2$~s, the nostril jet formed due to the fan located in the 'top' position containing a high level of CO$_2$ (36,000 ppm) was directed towards the bottom of the respirator. When the jet impinged on the interior surface of the fan, the jet dispersed with some of the gas moving towards the fan at the top-position. In the side-fan position, the redirection was toward the right-hand side of the mask where the fan was located, and this caused a circulation in the respirator. The fan's location at the bottom of the respirator is parallel to the nostril jets, which reduced the flow circulation inside the respirator. At $t=3.6$~s, the top-fan position is filled with high CO$_2$ concentration (>28,000 ppm), while the bottom-fan position maintained a lower CO$_2$ concentration during exhalation. 

Inhalation occurs from $t = 4 - 5.65$~s, and it is observed that during this period, the residual CO$_2$ inside the respirator during exhalation was discharged for the bottom-fan respirator. Therefore, fresh air containing a low level of CO$_2$ (385 ppm) entered the mask region, and high CO$_2$ concentration build-up is avoided during inhalation. However, for the top-fan position, high CO$_2$ concentration can be seen inside the respirator during inhalation.

\begin{figure}[h!]
	\centering
	\includegraphics[width=1.1\linewidth]{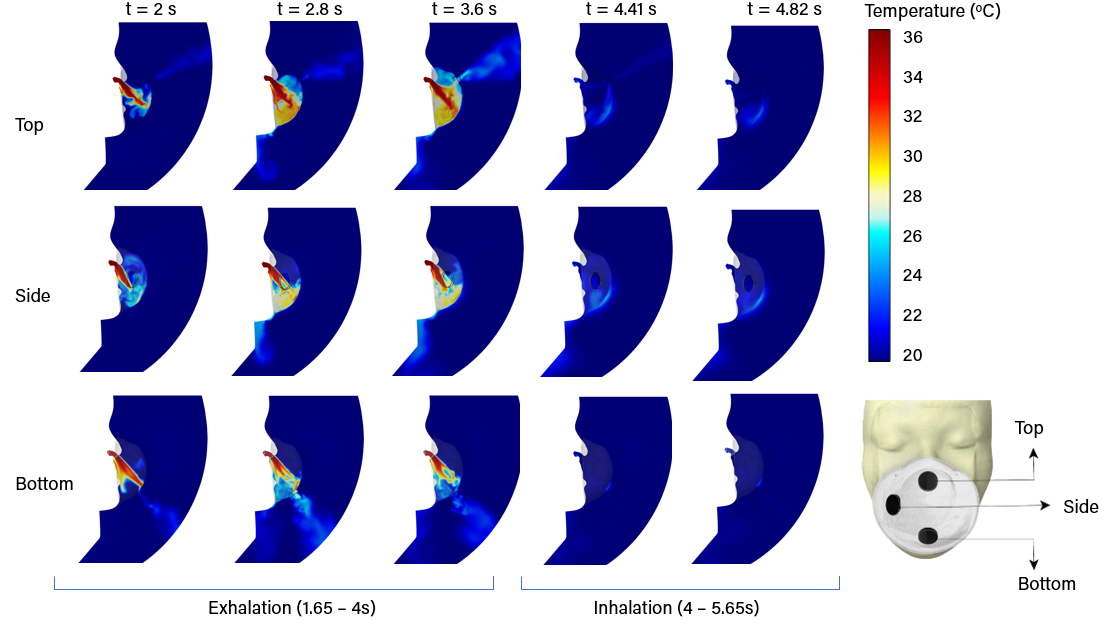}
	\caption{Air temperature distribution for the mid-sagittal plane of the right nasal passage for different fan locations}
	\label{fig:temp-con}
\end{figure}

Figure~\ref{fig:temp-con} illustrates the air temperature distribution during the first breathing cycle. During exhalation, the high-temperature air (36$^\circ$C) exited the nostrils and accumulated in the mask and the temperature inside the respirator increased. At $t=3.6$~s, the high-temperature air filled the entire top-fan position respirator compared to the other locations. For the bottom-fan position, the high-temperature air accumulated only around the bottom of the mask. During inhalation, the high-temperature air inside the respirator was inhaled back into the nasal cavity, and the temperature in the respirator decreased for different fan locations. At $t=4.4$~s the air temperature reached the ambient temperature (20$^\circ$C) for the bottom-fan respirator.
\clearpage

\begin{figure}[h!]
	\centering
	\includegraphics[width=1.1\linewidth]{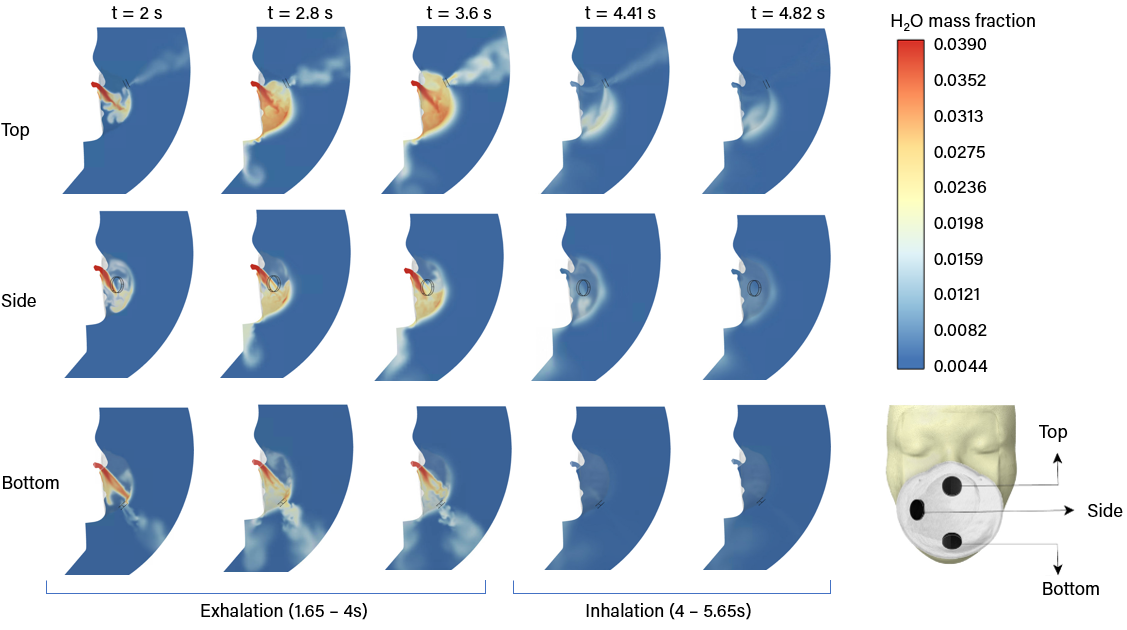}
	\caption{Water vapor mass fraction distribution at the mid-sagittal plane of the right nasal passage at different fan locations. The H$_2$O mass fraction is the specific humidity in kg of water vapour per kg of dry air. As a reference, the peak H$_2$O mass fraction of 0.039 provides 100\% relative humidity at 36$^\circ$C.}
	\label{fig:h2o-cont}
\end{figure}

\begin{figure}[h!]
	\centering
	\includegraphics[width=1.1\linewidth]{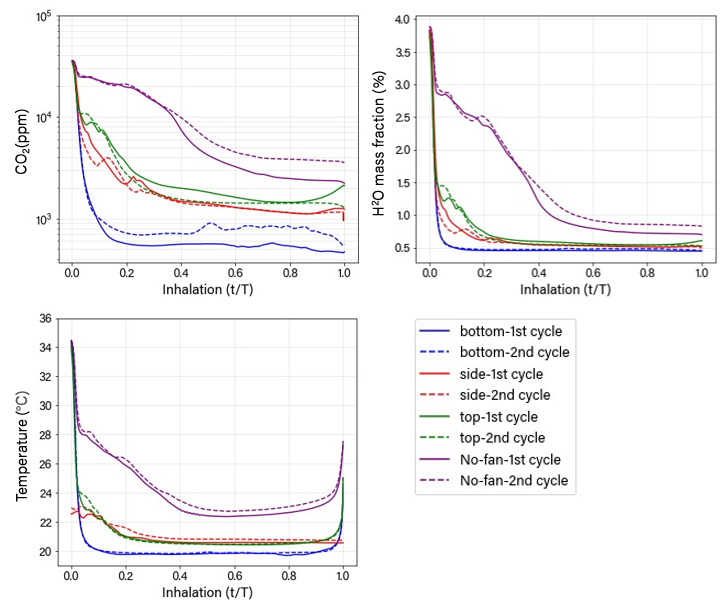}
	\caption{CO$_2$ concentration, water vapour mass fraction, and air temperature variation during the first and second breathing cycles. The results of the fan-powered respirator with different fan locations are compared with the no-fan respirator taken from \cite{salati2021n95}.}
	\label{fig:report}
\end{figure}

Figure~\ref{fig:h2o-cont} shows the water vapour mass fraction in the mid-sagittal plane of the right nasal nostril. The water vapour distribution follows that of the temperature distribution. The airflow from the lung during exhalation was 100\% humid, and the respirator trapped the moist, exhaled air, and the water vapour level within the respirator increased during exhalation. At $t=4.4$~s, the bottom-fan respirator expelled all the humid air and the water vapour mass fraction reached the ambient level for the inhalation.
\clearpage
Figure~\ref{fig:report} shows the  CO$_2$ concentration, water vapour mass fraction, and temperature during inhalation, and these results were compared with reported data of a respirator mask without a fan taken from \cite{salati2021n95}. Breathing without a fan respirator (labelled `no-fan') increased all parameters, i.e., an increase in the moisture and CO$_2$ content and temperature. The bottom-fan position produced the lowest inhaled CO$_2$ concentration (580 ppm) compared with other fan locations. A slight increase in the CO$_2$ level occurred in the second breathing cycle, which demonstrates the residual build-up of the gas-mixtures from exhalation. The bottom-fan position also produced the greatest temperature decrease to 20$^\circ$C. 

\section{Discussion}
N95 respirator masks are essential personal protective equipment (PPE) to filter inhaled air to protect individuals from environmental pollutants like viruses, pathogens, and allergens. Healthcare workers have the highest risk of exposure to COVID-19 virus, therefore must wear PPEs to ensure their safety. However, due to the high flow resistance and buildup of temperature and CO$_2$, while wearing the mask, healthcare workers prefer to use a less restrictive respirator while working in high-risk environments. Despite the compulsory regulations for the use of respirators within a healthcare environment \cite{goh2020face}, people refuse to wear respirators due to physical and breathing discomfort \cite{feng2020rational}. A modified N95 respirator can produce fresher air and improve breathing conditions. 

This study investigated the effect of different fan locations on an N95 respirator and analysed the airflow parameters of CO$_2$, temperature, and relative humidity inside the respirator. \cite{salati2021n95} demonstrated that respirator breathing increases the time-averaged inhaled air temperature by 7\% compared with normal breathing, which increases mucosal temperature, especially in the turbinate region, that reduced the air conditioning capacity of the nasal cavity. According to \citep{sullivan2014perception,zhao2011perceiving,lindemann2008impact}, nasal mucosal cooling (air conditioning function) affects the sensation and perception of ample nasal airflow. Figure \ref{fig:report} demonstrated that a fan-respirator decreased (i.e. cooled) the inhaled air temperature significantly where a bottom-fan position respiratory refreshed the air sufficiently to the ambient temperature of 20$^\circ$C. This finding demonstrates that using a fan-respirator eliminates the nasal fullness and stuffiness sensation, particularly when the fan is located at the bottom of the mask. A fan-less N95 respirator restricts mixing of the exhaled air with the ambient air, causing an accumulation of CO$_2$ inside this closed space area \citep{salati2021n95}. This study showed that adding a fan to the existing respirators can significantly reduce the accumulation of CO$_2$ inside the respirator. 

\begin{figure}[h!]
	\centering
	\includegraphics[width=1\linewidth]{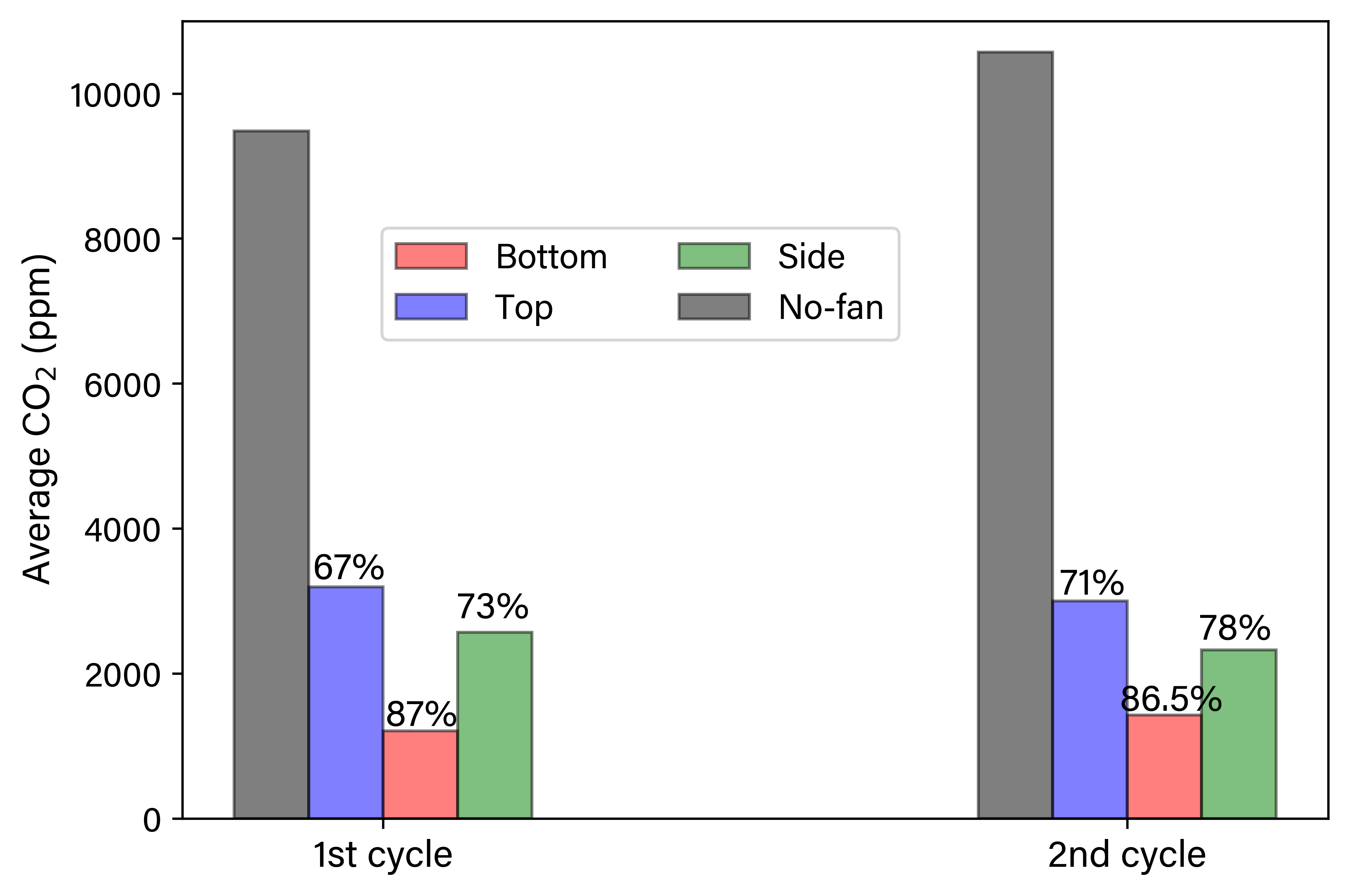}
	\caption{Time-averaged inhaled CO$_2$ using a fan-respirator for different fan locations. The results of the fan-powered respirator with different fan locations are compared with the no-fan respirator taken from \cite{salati2021n95}, and the reduction percentage are written on top of each bar.}
	\label{fig:co2average}
\end{figure}

Figure~\ref{fig:co2average} demonstrated the time-average inhaled CO$_2$ concentration rate during the first and second inhalation cycles of breathing in a respirator mask with the three positioned fans and without a fan. The data for the inhaled CO$_2$ concentration rate  without a fan was obtained from \cite{salati2021n95}. The effect of fan respirators was a significant decrease in CO$_2$ exposure as the fan allowed a portal for exhaled air to escape. The reduction percentage of the  CO$_2$ concentration for each fan position positioned respirator mask relative to the no-fan respirator is labelled above each bar in Figure~\ref{fig:co2average}. 
\clearpage
The bottom-fan respirator produced the greatest decrease in the average CO$_2$ concentration 87\% compared with the no-fan respirator breathing. The performance of the top- and side-fan positions also reduced the average CO$_2$ concentration but not as well as the bottom-fan position, where the reduction ranged from 67\% to 78\%. Including a fan in a respirator provided significant refresh of the air within the respirator by allowing the exhaled air to escape. The fans provided a reduction in the CO$_2$ concentration within the respirator to a non-hazardous level, thus allowing the user to wear the respirator for a prolonged period time safely, and avoiding the need to remove the respirator. 

Although our study investigated the effects of different fan locations in a scenario with a gap between the respirator and face, it is worth noting that fan respirators in a sealed respirator would be more beneficial in reducing CO$_2$ concentration, internal mask temperature, and humidity. Sealed respirators removes the additional gas exchange occurring through the gaps, thus there is a greater reliance on the fans to provide the gas exchange. Traditional respirators, such as N95 masks, have limitations in terms of breathability, comfort, and efficacy, particularly in high-risk environments with hazardous substances or infectious particles. These limitations can result in poor compliance with respiratory protection protocols, particularly in non-occupational settings where respiratory protection is not typically required. By contrast, the fan respirator offers improved ventilation and reduced CO$_2$ concentration, mask temperature, and humidity, which could increase wearer comfort and compliance with respiratory protection protocols. This could be particularly important in the context of the COVID-19 pandemic, where the use of respiratory protection has been widely recommended but may be met with resistance or non-compliance due to discomfort or other factors.

Furthermore, the presence of a large temperature difference between the ambient air at 20$^\circ$C and the human body at 34$^\circ$C, will induce a thermal plume that can influence the flow field in the breathing zone by countering the exhaled air to move upward. The presence of a thermal plume produces a vertical flow field where experimental and computational studies showed velocities of approximately 0.2 to 0.4 m/s \citep{tacutu2022experimental, zong2022review}. These low velocities are typical of natural convection flows that are much smaller than forced convection flows. \cite{salati2022exhaled} modelled exhalation flows incorporating a thermal plume, where despite the thermal plume the exhalation jet dominated the natural convective forces during most of the breathing period of the sinusoidal breathing waveform where both inhalation and exhalation are greater than 0.4 m/s. However, there is a small window of time during breathing when the flow is less than 0.4 m/s which occurs during the transition between inhalation and exhalation. In this period the thermal plume may have an influence on the flow field.

\section{Conclusion}
Inhaled gas-mixture properties, including temperature, moisture, and CO$_2$ were monitored for fan respirator breathing with different fan locations using CFD simulations. The results were compared with a no-fan respirator during breathing. The inhaled air was sufficiently refreshed when a bottom-fan position was used, where the air temperature in the mask for inhalation returned to the ambient temperature, eliminating the nasal stuffiness sensation associated with fan-less respirators. The average CO$_2$ concentration decreased significantly for all fan locations. The bottom-fan, side-fan, and top-fan positions reduced the average CO$_2$ concentration to 87\%, 73\%, and 67\%, respectively, compared with the no-fan respirator. The CO$_2$ level exposure using any fan-position respirator was within the non-hazardous level. The nostril jets during exhalation move towards the bottom of the mask. Therefore, the top-fan and side-fan respirators could not discharge the exhaled gas quickly. Overall, the bottom-fan respirator was more efficient in removing the nostril jet gas mixture with exhalation properties than the other fan respirators.

%\itemsep{0pt plus 0.1pt}
\small
\setstretch{1.0}
\clearpage
\bibliography{aircon.bib}

%===================================================
\end{document}